\documentstyle[sprocl]{article}

\input{psfig.sty}

\def\be{\begin{equation}}
\def\ee{\end{equation}}
\def\bea{\begin{eqnarray}}
\def\eea{\end{eqnarray}}
\def\ul{\underline}
\def\z0{\rm Z^0}
\newcommand{\as}{\alpha_{\rm s}}

\newcommand{\oaa}{{\cal O}(\as^2)}

\newcommand{\epem}{\rm e^+\rm e^-}

\newcommand{\amz}{\as(M_{\rm Z^0})}

\def\mz{M_{\rm Z^0}}

\def\d2{D_2}
\def\oq{\char'134}

\def\ecm{E_{cm}}

\def\m2{\mu^2}
\def\q{\rm q}

\def\p{\rm p}

\def\q2{Q^2}

\def\wamz{\overline{\as}(M_{\rm Z^0})}
\def\dwas{\Delta\overline{\as}}

\begin{document}

\title{JET PHYSICS AT LEP \\ AND \\ WORLD SUMMARY OF $\as$~\footnote{Presented
at the $IV^{th}$ Int. Symp. on Radiative Corrections, Barcelona, Sept. 8-12,
1998.}}

\author{S. BETHKE}

\address{III. Physikalisches Institut, RWTH \\ 
D - 52056 Aachen, Germany \\
e-mail: bethke@rwth-aachen.de}

\maketitle\abstracts{Recent results on jet physics and tests of QCD from
hadronic final states in $\epem$ annihilation at PETRA and at LEP
are reviewed, with special emphasis on hadronic event shapes, charged
particle production rates, properties of quark and gluon jets and
determinations of $\as$.
The data in the entire energy range from PETRA to
LEP-2 are in broad agreement with the QCD predictions.
The world summary of measurements of $\as$ is updated and a detailed
discussion of various methods to determine the overall error of
$\amz$ is presented.
The new world average is $\wamz = 0.119 \pm 0.004$. 
The size of the error depends on the treatment of correlated
uncertainties.
}

\section{$\epem$ Annihilation Data} \label{sec:data}

\begin{figure}[ht]
\psfig{figure=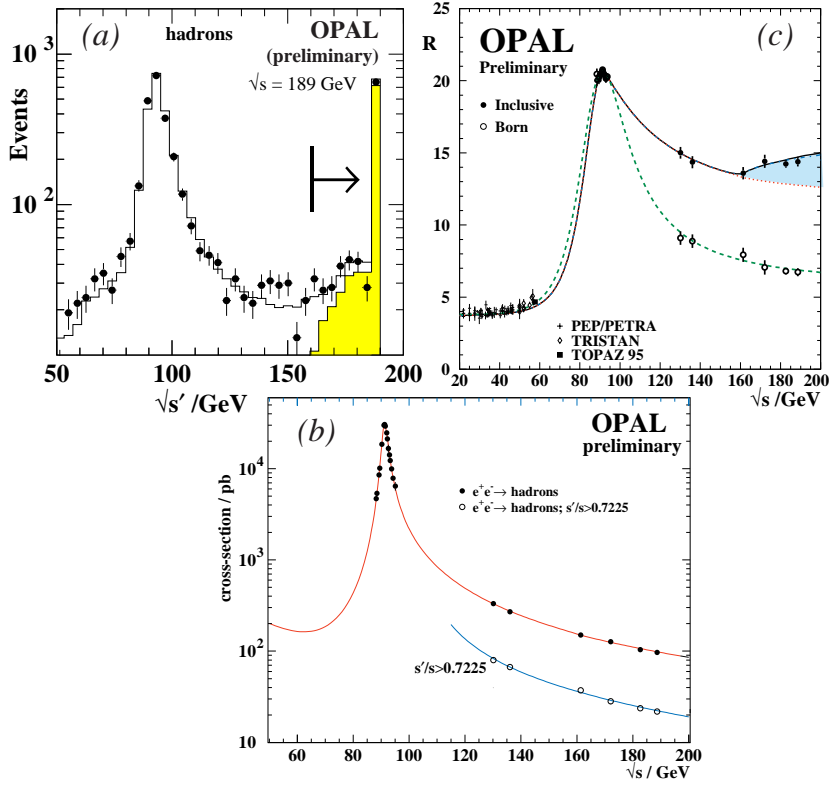,width=11.0cm}
\caption{Hadronic cross sections and available data in
$\epem$ annihilation (for details see text);
%$\sqrt{s}$: c.m. energy of hadronic system; 
R: = $\sigma(\epem\rightarrow q
\overline{q}) / \sigma(\epem \rightarrow \mu^+\mu^-)$.
\label{fig:sprime}}
\end{figure}

Over the past few years, a large amount of $\epem$
annihilation data in the c.~m. energy range from Q~$\equiv\sqrt{s} =$~10 to 189~GeV
was accumulated at the CESR, PETRA, PEP, Tristan, LEP and SLC 
accelerators.
The large data samples at LEP-1, which amount to about 4 million hadronic
events around the $\z0$ resonance for each of the four LEP experiments, and the most
recent data at the highest energies of LEP-2 (a few thousand events per experiment),
together with reanalysed PETRA data at lower c.m.
energies (about 50.000 hadronic events), provide powerful tools for precise tests of
perturbative QCD.

At energies below or at the $\z0$ resonance, respectively at PETRA and at LEP-1, the
study of
$\epem$ annihilation events is rather \oq easy" and straight forward: 
apart from two-photon processes, the energy and quantum numbers of the hard
scattering are well defined and different processes can be identified  and selected
with only very little backgrounds or biases.
At LEP-2, i.e. at energies above the $\z0$ pole, the situation is more
complicated:

\begin{itemize}
\item 
The annihilation cross section is orders of magnitude lower than at the
$\z0$ pole; see Figure~\ref{fig:sprime}b.
\item 
Initial state photon radiation reduces the available energy of the
hadronic c.m. system, $\sqrt{s'}$, see Figure~\ref{fig:sprime}a, and causes,
together with the resonant cross section around the $\z0$ mass, a large \oq
return-to-the-Z" effect.
Radiative events can be suppressed requiring a minimum reconstructed ratio of
$\sqrt{s'} / \sqrt{s}$ and other kinematic constraints.
\item 
Other processes like $\epem \rightarrow W^+W^-$ and $\epem \rightarrow \z0 \z0$
emerge above the respective energy thresholds, see the shaded area
in Figure~\ref{fig:sprime}c, causing
a certain irreducable background for QCD studies. 
\end{itemize}

In the following sections, recent QCD tests from LEP-1 ($\sqrt{s} \sim
91$~GeV; Section~2), from LEP-2 ($\sqrt{s} \sim 130$~GeV to 189~GeV; Section~3) and
from a combination of PETRA and LEP data ($\sqrt{s} \sim 14$~GeV to 172~GeV;
Section~4) are presented.
The world summary of $\as$ is updated in Section~5.

\section{QCD Tests at LEP-1} \label{lep1}

\subsection{$\as$ from Event Shapes using Optimised $\oaa$ QCD}

\begin{figure}[ht]
\psfig{figure=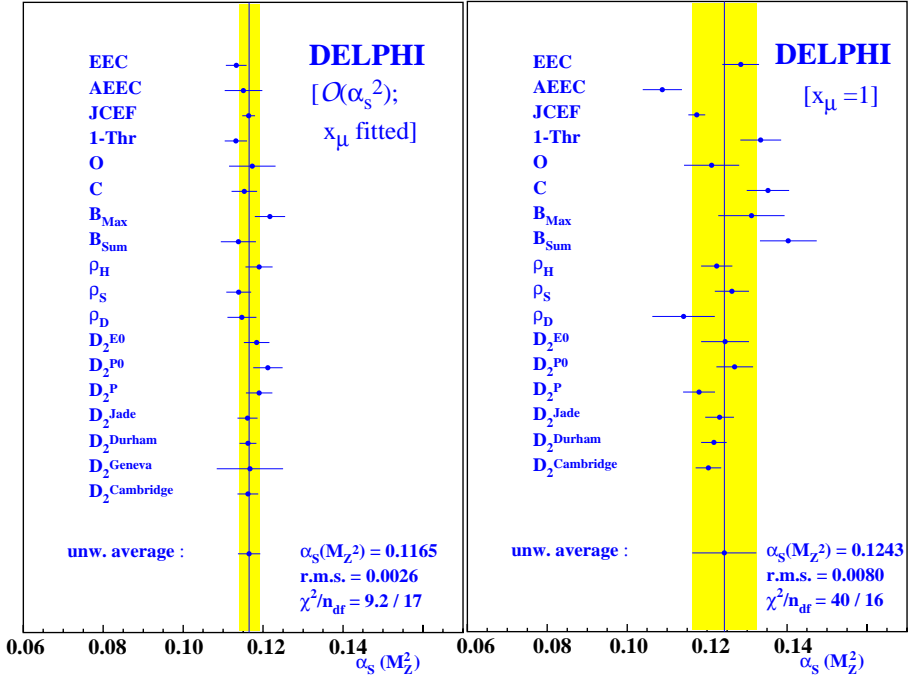,width=12.0cm}
\caption{Measurements of $\as$ from hadronic event shapes at
LEP-1 using $\oaa$ QCD predictions with (left) and without (right)
optimised renormalisation scales $\mu$.
\label{fig:d-as}}
\end{figure}

The DELPHI collaboration contributed a new measurement of $\as$ from oriented event
shape distributions at LEP-1~\cite{d-shape}.
17 different event shape observables are measured as a function of the polar angle
of the thrust axis, and $\as$ is determined from fits to $\oaa$ QCD calculations.
As already reported earlier~\cite{sbscale,d-as,o-as}, good agreement between
theory and data can be obtained if both $\as$ and the renormalisation scale $\mu$
are determined simultaneously; see Figure~\ref{fig:d-as}.

With optimised renormalisation scales and allowing for scale uncertainties between
$0.5\times\mu_{exp}$ and $2\times \mu_{exp}$, consistent results of
$\amz$ emerge, leading to a combined average of $\amz = 0.117 \pm 0.003$.
Both the average and the error, which includes theoretical uncertainties from
scale changes as given above, are smaller than those obtained from resummed $\oaa$
QCD fits. 
This is basically due to the choice of the averaging procedure and
error definition.\footnote{
Note that, for instance, a previous study~\cite{o-as} based on 13
observables and using a different procedure to average results and determine the
overall error obtained $\amz = 0.122^{+0.006}_{-0.005}$ which, if the same
procedure as used in the DELPHI analysis is applied, converts to $\amz = 0.116 \pm
0.003$.}
The broad consistency between data and optimised $\oaa$ QCD justifies
this procedure and suggests to reconsider optimised fixed order perturbation
theory as an alternative to resummation which was preferred in the past.

\subsection{Differences between Quark- and Gluon-Jets}

\begin{figure}[ht]
\psfig{figure=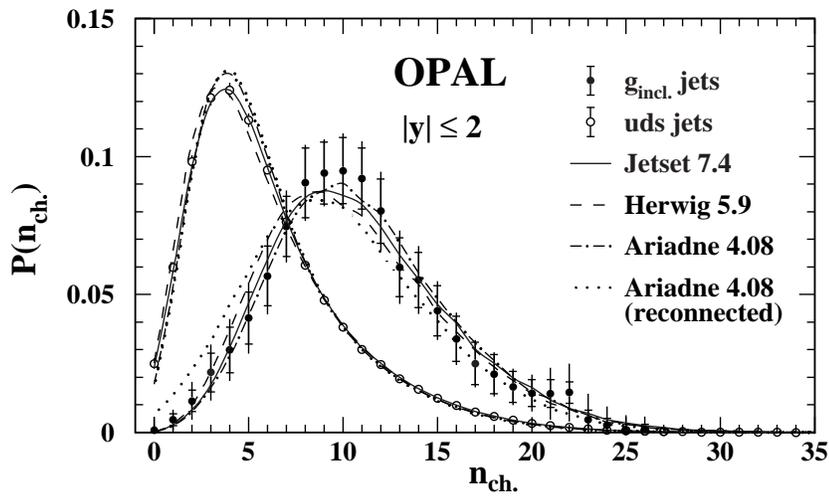,width=11.0cm}
\caption{Charged particle multiplicites in gluon-inclusive and in 
light (uds) quark jets in the central rapidity ($|y|<2$) range.
\label{fig:o-qgdiff}}
\end{figure}

\begin{figure}[ht]
\psfig{figure=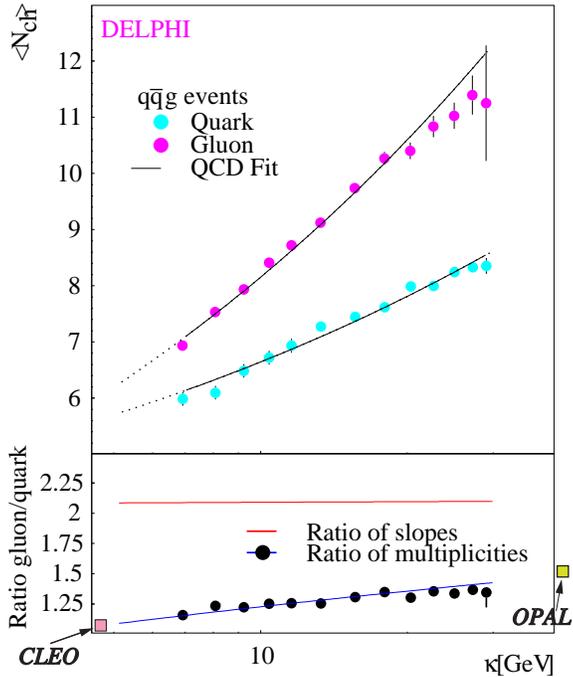,height=9.0cm}
\caption{Charged particle multiplicities in anti-tagged gluon and 
light quark jets as a function of the energy scale $\kappa$.
\label{fig:d-qgdiff}}
\end{figure}

Differences between quark- and gluon-jets were studied quite intensively during the
past few years.
The aim is to test the basic QCD prediction that hadrons coming from gluon
jets should exhibit a softer  energy spectrum and a wider transverse momentum
distribution than those originating from quark jets, due to the larger colour charge
of the gluon. 
In particular, the ratio $R_{qg}$ of the
average multiplicities of hadrons in gluon jets and in quark jets should, 
for infinite
jet energies and in leading order QCD, be $\approx C_A / C_F = 3 / (4/3) = 9/4 =
2.25$.

Experimental procedures to separate quark- from gluon-jets are usually based on
vertex tagging of primary b-quark decays. 
In short, 3-jet like events are selected in which one of the two lower energetic
jets is tagged as a b-quark, while the other low energy jet is then taken to be
the gluon jet.
From first analyses of this type it was found~\cite{o-qgdiff1} that, after correction
for misidentified jets,
$R_{qg} = 1.27 \pm 0.07$.
No QCD calculation for this particular type of analysis exists, such that a direct
comparison of this result with theory is not possible.

Theoretical predictions only exist for colour singlet $q\overline{q}$ and $gg$ final
states,  where however the latter state
is not experimentally accessible.
In order to perform an analysis closer to theory, 
OPAL followed a new strategy in which events 
with a high-energetic gluon-jet recoiling against a (vertex-tagged)
$q\overline{q}$ system~\cite{o-qgdiff2} were
selected. 
Such events are relatively rare, leading to about 550 selected gluon jets from 
OPAL's LEP-1 data sample. 

A comparison of the charged hadron multiplicity distribution of such gluon
hemispheres with those of ordinary light quark event hemispheres is shown in
Figure~\ref{fig:o-qgdiff}, where a significant difference between quark- and
gluon-jets is seen.
For a central rapidity range, the hadron multiplicity ratio $R_{qg}$ is found to
be $1.87 \pm 0.13$; the remaining difference to the QCD expectation of 2.25 is
likely to be explained by finite jet energy effects.

DELPHI has studied the scale dependence of particle multiplicities in quark- and
gluon-jets~\cite{d-qgdiff}.
Here, gluon-jets and light quark-jets are (anti-)tagged in 3-jet events, and a jet
energy scale of $\kappa = E_{jet} {\rm sin} (\theta /2)$, where $\theta$ is the
angle between the two lowest energetic jets, is defined.
The charged hadron multiplicities for quark- and gluon-jets, as a function of
$\kappa$, are diplayed in Figure~\ref{fig:d-qgdiff}.
Also shown is the ratio of multiplicities and the ratio of slopes; the latter
being close to the QCD prediction of 2.25.

From these measurements DELPHI determines the
ratio $C_A / C_F$ to be $2.27 \pm 0.012$ which is in good agreement with QCD, and in
particular with the expected colour charge of the gluon.

\section{QCD Tests at LEP-2} \label{sec:lep2}

\subsection{Hadronic Event Shapes and Running of $\as$}

\begin{figure}[ht]
\psfig{figure=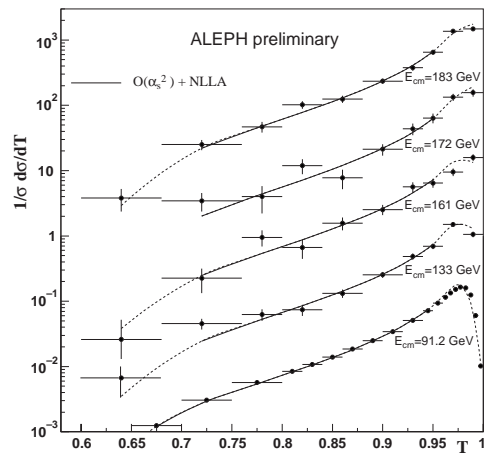,height=6.0cm}
\caption{Thrust-distributions measured by ALEPH at various
c.m. energies, and common fit to analytic QCD calculations in resummed 
next-to-leading order perturbation theory.
\label{fig:shapes}}
\end{figure}

\begin{figure}[ht]
\psfig{figure=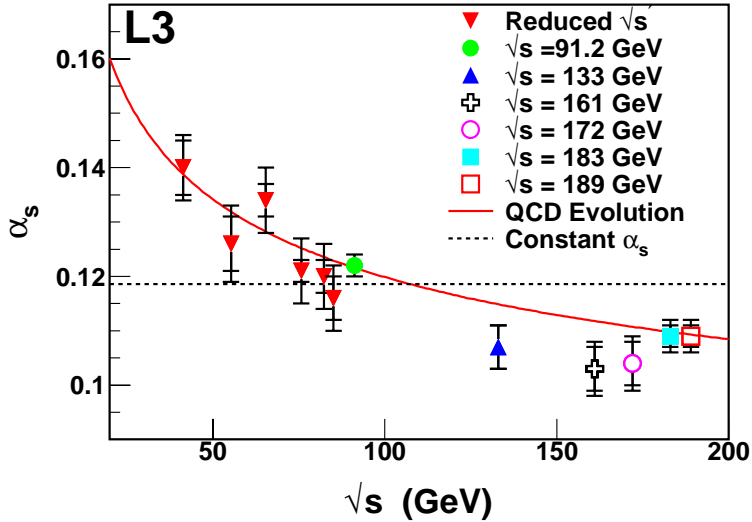,height=7.0cm}
\caption{Running of $\as$ as measured by L3.
\label{fig:L3-as}}
\end{figure}

At each new energy point of LEP-2, all four LEP experiments have extensively studied
hadronic event shape distributions and compared the new measurements with the
predictions of QCD Monte Carlo models, as well as with analytic QCD calculations.
In all cases, up to and including the most recent data at $\sqrt{s} = 189$~GeV, good
agreement of data and theory was found, and no significant deviation from the
standard expectation was seen.
As an example, Figure~\ref{fig:shapes} shows the thrust-distributions measured by
ALEPH~\cite{a-shapes} at LEP-1 and at four energy points of LEP-2, together with a
fit to resummed
$\oaa$ QCD calculations which is in good agreement with the data at all energies.

The L3 collaboration has summarised their measurements~\cite{l3-running} of $\as$
from various event shape distributions, at LEP-1, at all LEP-2 energy points, and at
hadronic c.m. energies below the $\z0$ pole, from an analysis of radiative events
recorded at LEP-1.
The data, which are displayed in Figure~\ref{fig:L3-as}, are in very good agreement
with the QCD prediction of a running coupling $\as (\sqrt{s})$.

\subsection{Energy Dependence of Charged Particle Production}

\begin{figure}[ht]
\psfig{figure=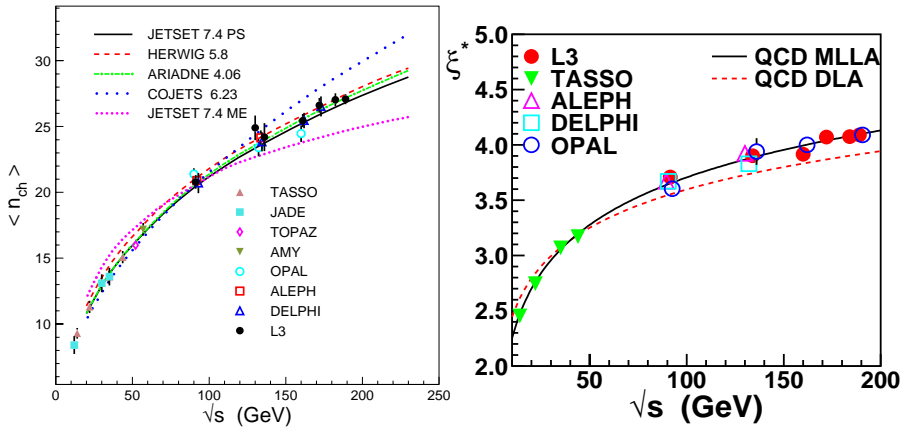,width=12.0cm}
\caption{Average charged particle multiplicities and peak position 
$\xi^*$ of
the $\xi = {\rm ln}(1/x)$ distribution (compilation by L3).
\label{fig:multiplicities}}
\end{figure}

The energy dependence of particle production was, similarly as hadronic event shapes,
continuously monitored by all LEP experiments.
The variation of such observables with energy is found to be in good agreement
with QCD predictions, and also with the \oq standard" QCD plus hadronisation models.
The energy
dependence of the average charged hadron multiplicity and of the peak position
$\xi^*$ of the $\xi = {\rm ln}(1/x)$ distribution ($x = p /
E_{beam}$)~\cite{l3-running} are shown in Figure~\ref{fig:multiplicities}.

\section{Power Corrections and Energy Dependence of Event Shapes}
\label{sec:powcor}

\begin{figure}[ht]
\psfig{figure=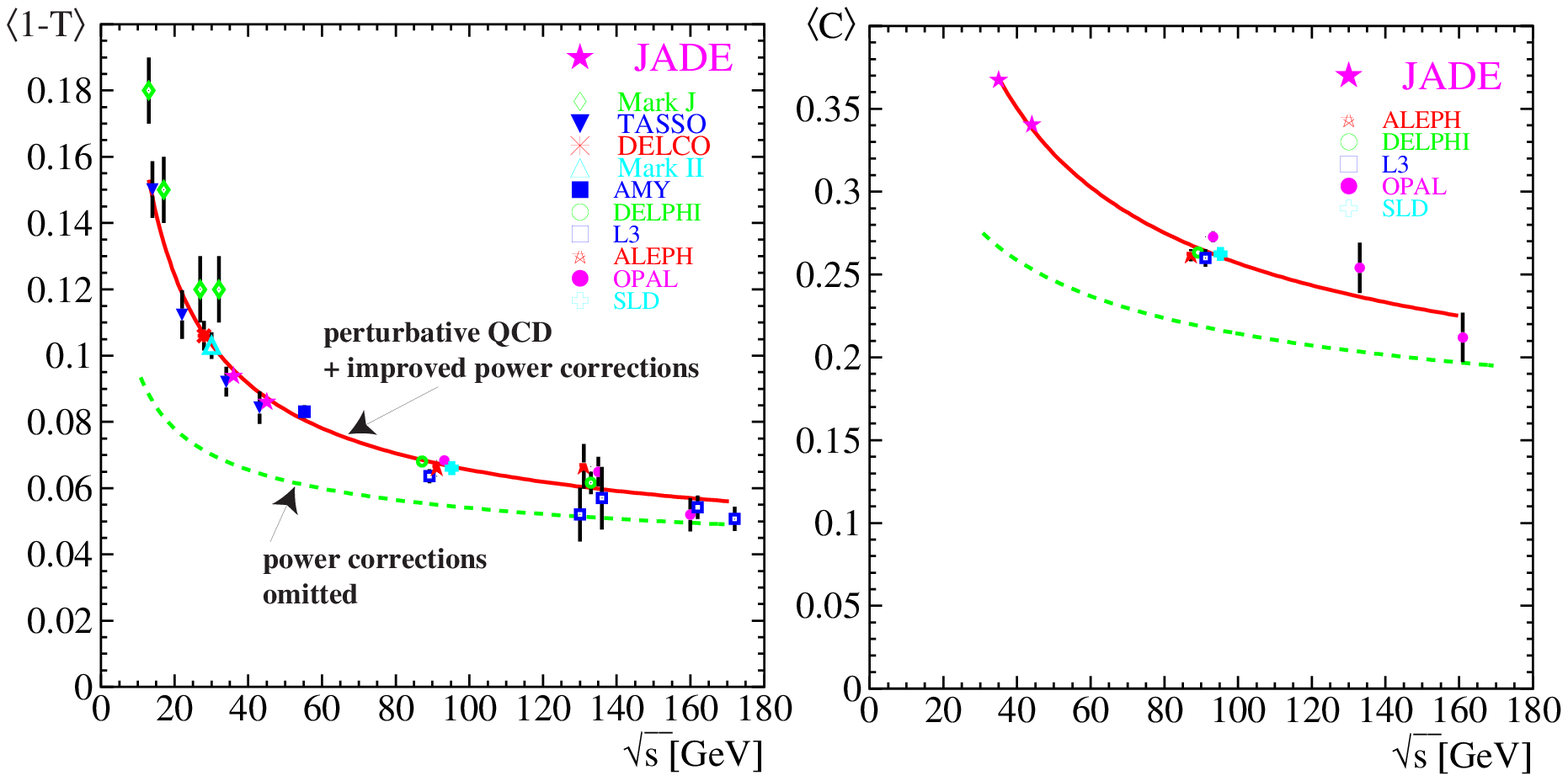,width=12.0cm}
\caption{Energy dependence of mean values of (1-Thrust) and of the 
C-parameter.
\label{fig:jade-tc}}
\end{figure}

\begin{figure}[ht]
\psfig{figure=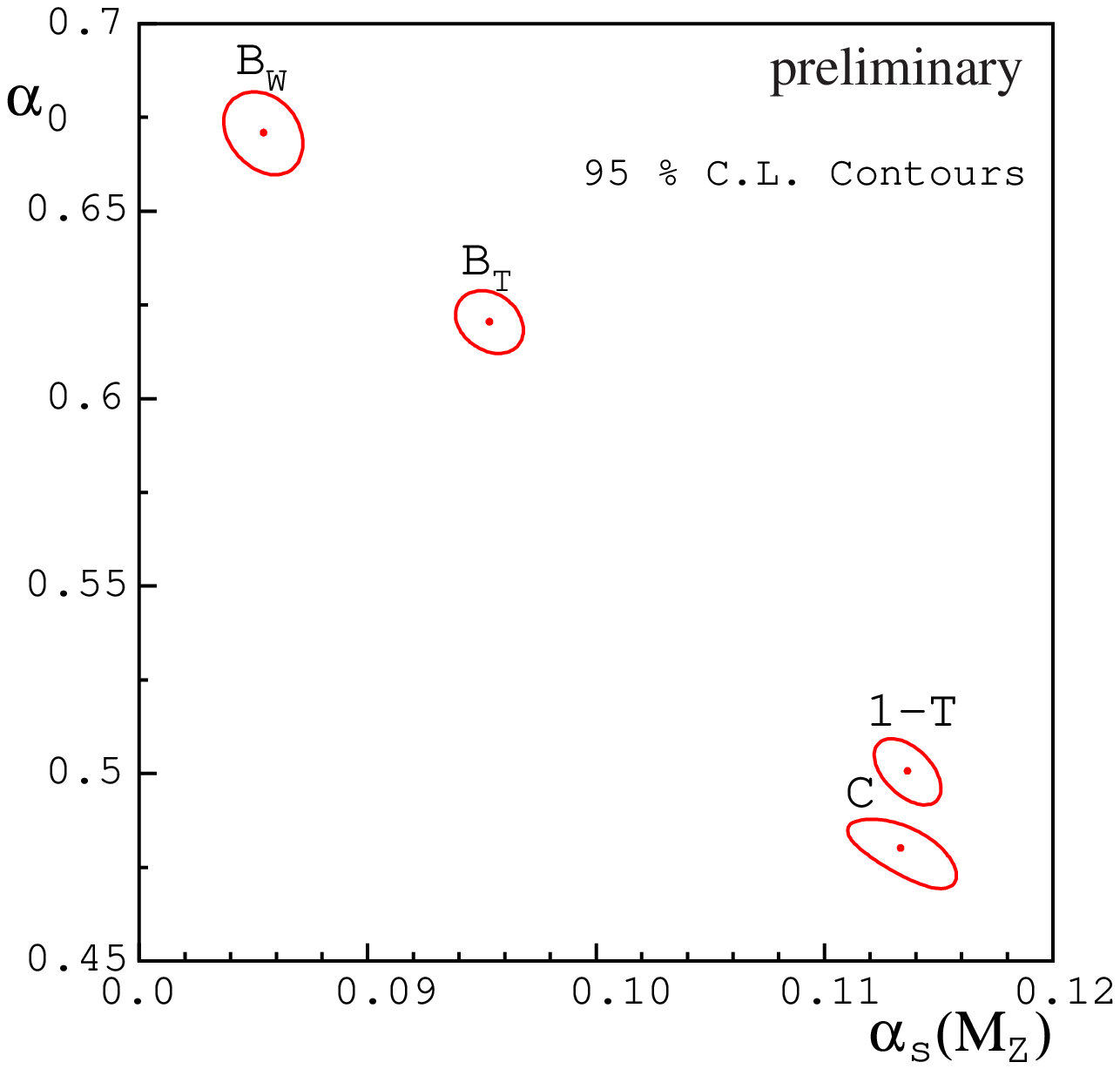,height=6.0cm}
\caption{\label{fig:jade-a0as}
Fit results for $\alpha_0$ and $\amz$ for several differential
event shape distributions measured 
in the c.m. energy range from 35 to 183 GeV.}
\end{figure}

The energy dependence of mean values of event shape observables can be
parametrised by the $\oaa$ perturbative QCD prediction \cite{ert} plus a term
including the improved two-loop calculations (the \oq Milan factor") of
power-suppressed 1/Q non-perturbative contributions~\cite{powcor1}, the so-called \oq
power corrections".
The latter contains the moment $\alpha_0$ of an effective coupling below an infrared
scale $\mu_I$, which is expected to be universal for all applicable event shape
observables.

A compilation of available data on the mean value of (1-thrust) and of the
C-parameter is shown in Figure~\ref{fig:jade-tc}, which is taken from a recent
re-analysis of JADE data at PETRA energies~\cite{jade2}.
Perturbative QCD plus power corrections is found to give a very good description of
the data, with $\amz = 0.118 \pm 0.002 \pm 0.004$ as determined from these data.
However, universality of $\alpha_0$ is only found to be satisfied at a level of 30\%.

For {\em differential} event shape distributions, the power corrections are
simply a shift of the perturbative ($\oaa$) spectra, and these were also studied
in a wide c.m. energy range, including the most recent PETRA and LEP data for a total
of four event shape observables~\cite{fernandez}.
A fit to these data, with $\amz$ and $\alpha_0$ as free
parameters for each observable, leads to the results displayed in
Figure~\ref{fig:jade-a0as}. 
Agreement in both $\amz$ and
$\alpha_0$ is obtained, to a good level of accuracy, for two of the
observables. 
However, the jet broadening parameters
$B_w$ and $B_t$ deviate significantly~\cite{fernandez} in both $\amz$ and
$\alpha_0$.

Most recently, the reason for these deviations was identified and traced
to the theoretical predictions~\cite{yuri}; a cure of this problem should soon be
available.

\section{World Summary of $\as$} \label{sec:as}

Significant determinations of the strong coupling strength, $\as$, remain to be
a demanding and interesting topic in experimental as well as
theoretical study projects in high energy physics.
In the following subsections, previous summaries of $\as$ measurements
\cite{qcd96,qcd97} will be updated and a new world average 
$\wamz$ will be determined.
Instead of a complete reference to all available measurements,
only the newest results are briefly introduced, and more
emphasis is spent on a detailed discussion of the overall {\em uncertainty} of
$\wamz$, $\dwas$.

\subsection{Updates and New Results}

The results of all significant determinations of $\as$, i.e. of all those which
are based on QCD calculations which are complete - at least - to next-to-leading
order perturbation theory, are summarised in Table~\ref{astab}.
The following entries were added or updated since summer 1997 \cite{qcd97}
(underlined in Table~\ref{astab}):

\begin{itemize}
\item
The most recent determination of $\as$ from the GLS sum rules, based on
new data from $\nu -N$ scattering~\cite{ccfr-gls} is included, replacing 
the previous result from Chyla and Kataev~\cite{kataev}.
\item
New measurements of $\as$ from high statistics studies of  vector
and axial-vector spectral functions of hadronic $\tau$-decays are available
from ALEPH~\cite{a-tau} and OPAL~\cite{o-tau}.
As the most complete and precise studies of $\tau$ decays to date, these
results are combined and taken to replace earlier results~\cite{old-tau}.
\item
H1 has contributed new determinations of $\as$ from
(2+1)-jet event rates at HERA \cite{h1-jet-new}, replacing a previous measurement
\cite{h1-jet}.
A combination of these new results with a former one from ZEUS
\cite{zeus-jet} is updated in Table~\ref{astab}.
\item
A new determination of $\as$ from $\Upsilon$-decays
\cite{jamin-pich} replaces  earlier results \cite{kobel}.
\item
A recent determination of $\as$ from the total $\epem$ hadronic cross section
measured by CLEO at $\ecm = 10.52$~GeV \cite{cleo-rhad} is added.
\item
Determinations of $\as$ from JADE data, at $\ecm=35$ and 44~GeV, were
updated \cite{jade2} by the inclusion of another observable, the
C-parameter.
\item
$\as$ from the most recent LEP result on
$R_l = \frac{\Gamma (Z^0 \rightarrow hadrons)}{\Gamma (Z^0 \rightarrow
leptons)}$  was updated~\cite{lep-ew}
(these results are still preliminary).
\item
LEP results on $\as$ from event shapes measured at $\ecm = 183$ and
189~GeV~\cite{lep-shapes}
are combined and added to the list (some of these results are still preliminary).
\end{itemize}

\begin{figure}[ht]
\psfig{figure=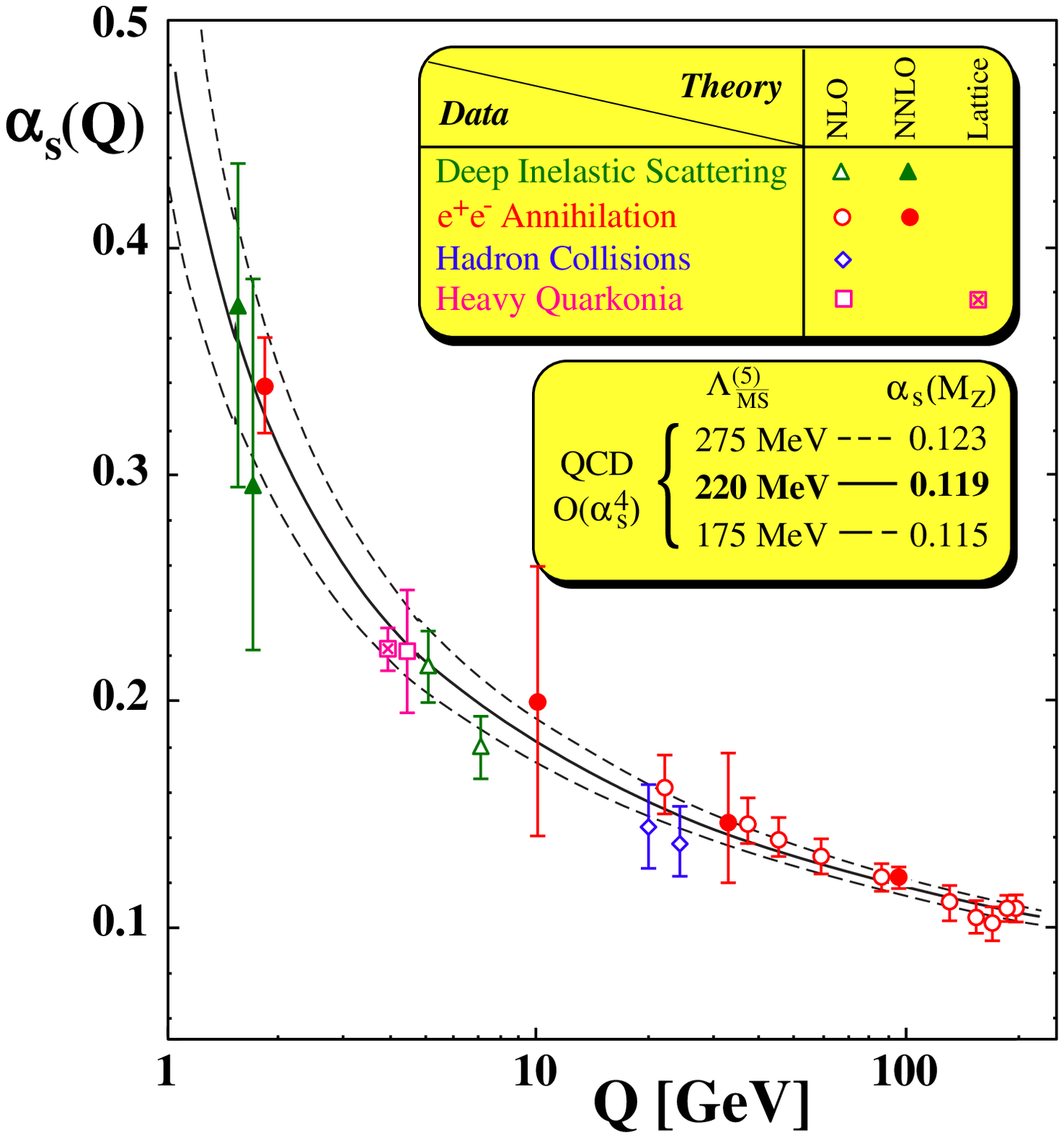,height=10.0cm}
\caption{Summary of $\as (Q)$ .
\label{fig:as-q}}
\end{figure}

Further interesting and recent developments are a QCD analysis of neutrino
deep inelastic scattering data for $xF_3$, in next-next-to-leading order of
perturbation theory (NNLO)~\cite{kataev-xf3}, resulting in $\amz = 0.118 \pm 0.006$,
and a reanalysis of muon deep
inelastic scattering data~\cite{newmudis}, resulting in $\amz = 0.118 \pm 0.002
\ (stat+syst)$.
Both these results are subject to further completion and verification; they are
therefore considered to be preliminary and are not included in this summary.

\begin{figure}[ht]
\psfig{figure=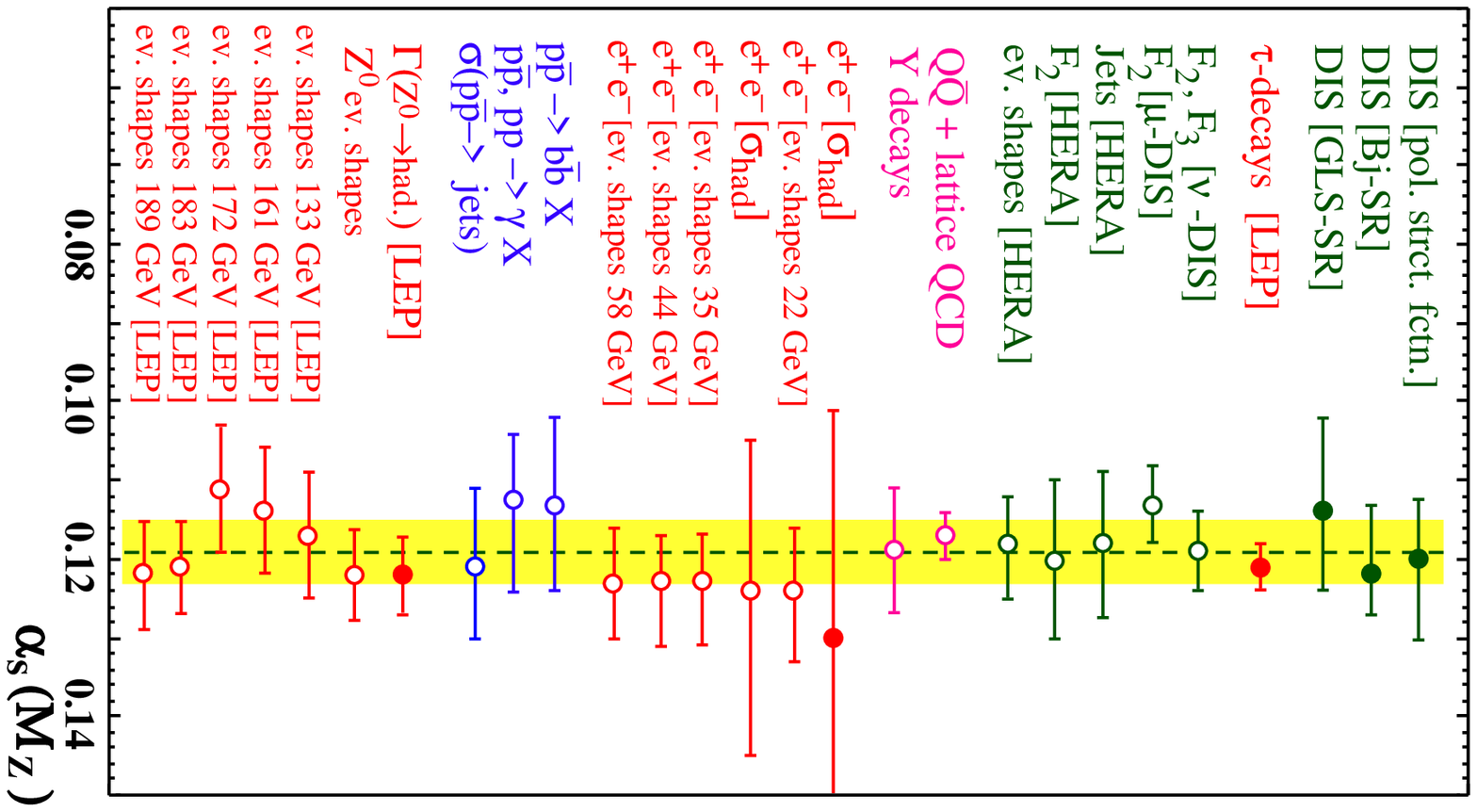,width=12.0cm}
\caption{Summary of $\amz$ .
\label{fig:as-mz}}
\end{figure}

\begin{figure}[ht]
\psfig{figure=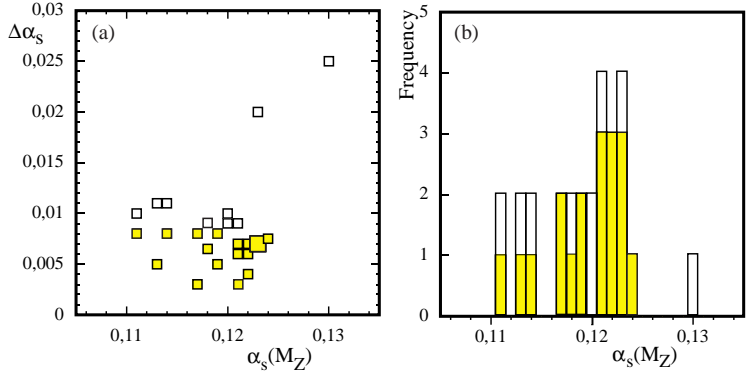,height=5.0cm}
\caption{Distribution of results of $\amz$ .
Shaded entries are those which are included in the final calculation of $\wamz$ and
its overall error; see text for details. 
\label{fig:as-hist}}
\end{figure}

The results for $\as (Q)$, given in the $3^{rd}$ row of Table~\ref{astab},  are
presented in Fig.~\ref{fig:as-q}. 
These results are evolved
from the energy scale $Q$, i.e. the typical energy scale of the hard scattering
process under study, to the reference energy scale
$\mz$, by using the QCD 4-loop beta-function with 3-loop matching at quark pole
masses
$M_b = 4.7$~GeV and
$M_c = 1.5$~GeV \cite{4-loop}, resulting in the values of $\amz$ given in the
$4^{th}$ row of Table~1.
These values of $\amz$ are displayed in Fig.~\ref{fig:as-mz}.
The distribution of all $\amz$ results is shown in a scatter plot of $\amz$ versus
its quoted error,
$\Delta\as$, and in a frequency distribution of $\amz$ 
(Figure~\ref{fig:as-hist}).

The world average value of $\amz$ and its overall  uncertainty as well as the
corresponding QCD curves shown in Figs.~\ref{fig:as-q} and~\ref{fig:as-mz} will be
discussed in the following section.

\subsection{World Average and Overall Uncertainty of $\amz$}

In order to
average the results of $\amz$, a weigthed mean of the quoted central values is
calculated, for all results as well as for subsamples as listed in
Table~\ref{aserr}. The weight of a measurement is taken to be the inverse of the
square of its total error.

The central value $\wamz$ does not depend on details of the weighting method; no
significant differences are found if, for instance, the simple and
unweighted mean is taken. 
Also, and more important, there are no significant differences between values of
$\wamz$ calculated from different subsamples of the data:
the averages from the high- and the low-energy data as well as those from $\epem$
annihilation, from DIS and from $p\overline{p}$ colliders agree well with each
other and with the overall average of $\wamz$ = 0.119, within the respective errors
(irrespective of how those errors are defined; see the discussion below).

While the central value of $\wamz$ is remarkably stable and well defined, the
determination of its overall uncertainty, $\dwas$,
depends very much on the detailed
definition of what this uncertainty should be, and how it should be calculated.
There are several reasons for this situation:

The errors of most $\as$ results are dominated by theoretical uncertainties,
which are estimated using a variety of different methods and definitions.
The significance of these non-gaussian errors is largely unknown.
Furthermore, there are large correlations between different results, due to
common theoretical uncertainties, as e.g. in the case of various event shape
measurements in
$\epem$ annihilations.
Nothing is known, however, about possible correlations between
$\as$ determinations from different processes, such as DIS and $\epem$ annihilations,
or between different procedures and observables used within the same class of
processes.

Therefore, in the past, the value of $\dwas$ was often \oq guess-timated",
and/or a variety of mathematical methods was applied to obtain a reasonable
estimate.
Some of these methods
will be applied and discussed in the following; the results are summarised in
Table~\ref{aserr}:

%%%%%%%%%%%%%%%%%%%%%%%%%%%%%%%%%%%%%% start table aserr
\renewcommand{\arraystretch}{1.3}
\begin{table}[htb]
\caption{
Average values of $\wamz$ plus averaged uncertainties, for several
methods to estimate the latter, and for several subsamples 
of the available data. \label{aserr} }
\begin{center}
  {\tiny 
\begin{tabular}{|l|c|c|c|c||c|c|}
   \hline
 & & uncorrel. & simple rms &  rms box & opt. corr. & overall \\
sample \hfill (entries)& $\wamz$ & $\dwas$ & 
  $\dwas$ & $\dwas$ 
  & $\dwas$ & correl.\\
\hline
all \hfill (27)         & 0.1193 & 0.0012 & 0.0044 & 0.0059 & 0.0049 & 0.71\\
 & & & & & &\\
$\dwas\le 0.008$ \hfill (18)
                        & 0.1193 & 0.0013 & 0.0038 & 0.0052 & 0.0042 & 0.64\\
$\dwas\le 0.006$ \hfill (7)
                        & 0.1190 & 0.0016 & 0.0033 & 0.0041 & 0.0030 & 0.49\\
$\dwas\le 0.004$ \hfill (2)
                        & 0.1190 & 0.0021 & 0.0028 & 0.0028 & 0.0022 & 0.11\\
 & & & & & &\\
only $\epem$\hfill (15) & 0.1210 & 0.0016 & 0.0045 & 0.0059 & 0.0052 & 0.77\\
only DIS\hfill (8)      & 0.1175 & 0.0025 & 0.0029 & 0.0053 & 0.0061 & 0.80\\
only $p\overline{p}$  \hfill (3)  
                        & 0.1156 & 0.0057 & 0.0053 & 0.0072 & 0.0088 & 0.69\\
 & & & & & &\\
$Q \le 10$~GeV  \hfill (9)
                        & 0.1184 & 0.0016 & 0.0029 & 0.0045 & 0.0038 & 0.69\\
$Q \ge 30$~GeV  \hfill (14) 
                        & 0.1199 & 0.0020 & 0.0047 & 0.0062 & 0.0060 & 0.69\\
\hline
\end{tabular} }
\end{center} 
\end{table}

\begin{itemize}
\item
For illustrational purposes only, an overall error is calculated assuming that all
measurements are entirely uncorrelated and all quoted errors are gaussian.
The results are displayed in the $3^{rd}$ column of Table~\ref{aserr}.
\item
The simple, unweigthed $r.m.s.$ of the mean value of all measurements is calculated 
and shown in the $4^{th}$ column, labelled \oq simple rms".
\item
Assuming that each result of $\amz$ has a rectangular-shaped rather than a
gaussian probability
distribution, 
all resulting weights (the inverse of the square of the total
error) are summed up in a histogram, and the resulting $r.m.s.$ of that distribution
is quoted as \oq rms box" \cite{qcd97}.
\item
A correlated error is calculated from the error covariance matrix, assuming
an overall correlation factor between all measurements.
The correlation factor is adjusted such that the total $\chi^2$ is one per degree
of freedom~\cite{schmelling}.
The resulting errors and 
correlation factors are given in the last two columns of
Table~\ref{aserr} (labelled \oq optimised correlation").
\end{itemize}

All of the methods defined above have certain advantages but also obey inherent
problems.
The \oq simple rms" indicates the scatter of all results around
their common mean, but does not depend on the individual errors
quoted for each measurement. 
The \oq box rms", which takes account of the individual errors
and of their non-gaussian nature, was criticised to be
too conservative an estimate of the overall uncertainty of $\as$. 
The \oq optimised correlation" method is closest to a mathematically appropriate
treatment of correlated errors, however - in the absence of a detailed knowledge of
these correlations - over-simplifies by the (unphysical ?) assumption of one
overall correlation factor, identical to all pairs of measurements. 
Moreover, the
$\chi^2$ calculated from the covariance matrix does not have, if correlations are
present, the same mathematical and propabilistic meaning as in the
case of uncorrelated data. 
In the extreme, $\chi^2$ may even be negative.

With these reservations in mind, all three methods do provide some  estimate of 
$\dwas$. 
Apart from systematic differences in the size of $\dwas$, they all depend on
the significance of the data included in the averaging process: in all cases,
$\dwas$ is largest if all data are included, and
tend to smaller values if the averaging is restricted to results
with errors $\Delta\as \le \Delta\as^{(max)}$, i.e. if only the most
significant results are used to calculate $\wamz$ and $\dwas$.

\begin{figure}[ht]
\psfig{figure=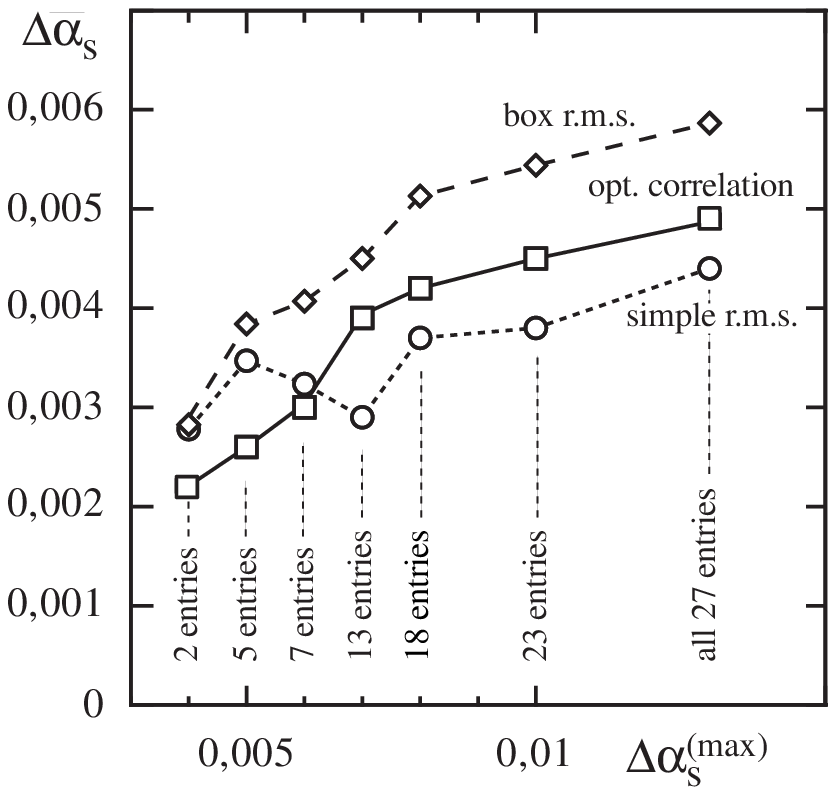,height=5.0cm}
\caption{Dependence of $\dwas$ on the
selection of results with errors $\Delta\amz \le \Delta\as^{(max)}$. 
\label{fig:as-err}}
\end{figure}

This can be seen from Table~\ref{aserr}, where $\wamz$ and $\dwas$ are
also shown for three subsets of data with $\Delta\as \le$ 0.008, 0.006 and 0.004.
The decrease of $\dwas$ as a function of $\Delta\as^{(max)}$ is graphically
shown in Figure~\ref{fig:as-err}.
All three estimates of $\dwas$, the \oq simple rms", the \oq box rms" and the
\oq optimised correlation", decrease from intial values of 0.004 ... 0.006,
if all $\as$ results are taken into account, to about 0.003 if only the most \oq
precise" results are included.
Only at the very extreme, taking the two results with the smallest quoted errors, the
\oq optimised correlation" method yields an overall error of less than 0.003.
Note that, despite the dependence of $\dwas$ on the choice of
$\Delta\as^{(max)}$, the overall average $\wamz$ does $not$ depend on this
selection!

On first sight it seems logical to restrict the determination of $\wamz$, and
especially of $\dwas$, to the most significant data, if inclusion of 
insignificant measurements enlarges $\dwas$.
Taken to the extreme, one may even be tempted to quote the one result which carries
the smallest quoted error as the final world average value of $\amz$ and
$\dwas$.
However, the errors on $\amz$ estimated in individual
studies are, in general, {\em lower limits} because unknown
and additional systematic effects can only increase the total error.
Small systematic errors of single measurements may well be due to ignorance,
over-optimism and/or neglection of certain error sources, which may be
difficult to judge.
Indeed, the errors of the two results with the smallest errors quoted, $\as$ from
$\tau$-decays and from heavy quark bound states using lattice gauge theory, were
often criticised as being overly optimistic.

In this sense, averaging over a number of well understood and
commonly accepted measurements of reasonable precision is a safe basis to
estimate $\wamz$ and $\dwas$.
While there is still a large degree of flexibility to choose the final data set and
the procedure to estimate $\dwas$, the choice to select results with 
$\Delta\amz \le 0.008$ and to determine $\dwas$ using the \oq optimised
correlation" method \cite{schmelling} seems reasonable and to be neither overly
optimistic nor pessimistic. The final world average of $\amz$ is therefore quoted to
be, c.f. Table~\ref{aserr} and Figure~\ref{fig:as-err},
$$ \wamz = 0.119 \pm 0.004\ . $$
The central values of 19 out of the 27 measurements
listed in Table~\ref{astab}, or equivalently 70\%, are inside this error 
range of $\pm 0.004$,
thus being compatible with the expectation of a \oq one standard deviation" interval.

If the result of $\as$ which is based on lattice gauge theory~\cite{lgt} is omitted
in the averaging process,
$\wamz$ and $\dwas$ increase
to $0.120 \pm 0.005$.
The same is true if the result
of the reanalysis~\cite{newmudis} of muon deep
inelastic scattering data is used\footnote{
The increase of $\dwas$ is due to an artifact of the \oq optimised correlation"
method which may increase the overall correlation factor (and thus, the overall
error) if individual measurements are closer to their common mean.} instead of the
one~\cite{virchaux} listed in Table~\ref{astab}.

Note that the value of $\dwas$ = 0.004 is a
factor of two larger than the one quoted in the latest edition
of the Review of Particle Physics~\cite{pdg98}. 
The smaller value of 0.002 quoted there corresponds
to a (slighly enlarged) r.m.s. assuming all measurements to be {\em totally
uncorrelated}; an assumption which seems, in view of the results discussed above,
unrealistic.

%\section{Summary} \label{sec:summary}

\section*{Acknowledgments}
I am grateful to W. Bernreuther, to O. Biebel, to S. Catani, to A. Kataev and to Yu.
Dokshitzer for helpful discussions.

%%%%%%%%%%%%%%%%%%%%%%%%%%%%%%%%%%% Begin Table astab
\renewcommand{\arraystretch}{1.7}
\begin{table}[h]
{\tiny
\caption{
World summary of measurements of $\as$.
Underlined entries 
are new or updated since summer 1997
(DIS = deep inelastic scattering; GLS-SR = Gross-Llewellyn-Smith sum rules;
Bj-SR = Bjorken sum rules;
(N)NLO = (next-)next-to-leading order perturbation theory;
LGT = lattice gauge theory;
resum. = resummed next-to-leading order). \label{astab}}
\begin{center}
\begin{tabular}{|l|c|l|l|c c|c|}
   \hline 
  & Q & & &  \multicolumn{2}{c|}
{$\Delta \amz $} &  \\ %\cline{5-6}
Process & [GeV] & $\alpha_s(Q)$ &
  $ \amz$ & exp. & theor. & Theory \\
\hline \hline %\normalsize
DIS [pol. strct. fctn.] & 0.7 - 8 & & $0.120\ ^{+\ 0.010}
  _{-\ 0.008}$ & $^{+0.004}_{-0.005}$ & $^{+0.009}_{-0.006}$ & NLO \\
DIS [Bj-SR] & 1.58
  & $0.375\ ^{+\ 0.062}_{-\ 0.081}$ & $0.121\ ^{+\ 0.005}_{-\ 0.009}$ & 
  -- & -- & NNLO \\
\ul{DIS [GLS-SR]} & 1.73
  & $0.295\ ^{+\ 0.092}_{-\ 0.073}$ & $0.114\ ^{+\ 0.010}_{-\ 0.012}$ & 
  $^{+0.005}_{-0.006}$ & $^{+0.009}_{-0.010}$ & NNLO \\
\ul{$\tau$-decays} 
  & 1.78 & $0.339 \pm 0.021$ & $0.121 \pm 0.003$
  & 0.001 &  0.003 & NNLO \\
DIS [$\nu$; ${\rm F_2\ and\ F_3}$]  & 5.0
  & $0.215 \pm 0.016$
   & $0.119\pm 0.005$   &
    $ 0.002 $ & $ 0.004$ & NLO \\
DIS [$\mu$; ${\rm F_2}$]
     & 7.1 & $0.180 \pm 0.014$ & $0.113 \pm 0.005$ & $ 0.003$ &
     $ 0.004$ & NLO \\
DIS [HERA; ${\rm F_2}$]
     & 2 - 10 &  & $0.120 \pm 0.010$ & $ 0.005$ &
     $ 0.009$ & NLO \\
\ul{DIS [HERA; jets]}
     & 10 - 100 &  & $0.118 \pm 0.009$ & $ 0.003$ &
     $ 0.008$ & NLO \\
DIS [HERA; ev.shps.]
     & 7 - 100 &  & $0.118\ ^{+\ 0.007}_{-\ 0.006}$ & $ 0.001$ &
     $^{+0.007}_{-0.006}$ & NLO \\
${\rm Q\overline{Q}}$ states
     & 4.1 & $0.223 \pm 0.009$ & $0.117 \pm 0.003 $ & 0.000 & 0.003
     & LGT \\
\ul{$\Upsilon$ decays}
     & 4.13 & $0.220 \pm 0.027$ & $0.119 \pm 0.008
     $ & 0.001 & $0.008$ & NLO \\
\ul{$\epem$ [$\sigma_{\rm had}$] }
     & 10.52 & $0.20\ \pm 0.06 $ & $0.130\ ^{+\ 0.021\ }_{-\ 0.029\ }$
     & $\ ^{+\ 0.021\ }_{-\ 0.029\ }$ & -- & NNLO \\
$\epem$ [ev. shapes]  & 22.0 & $0.161\ ^{+\ 0.016}_{-\ 0.011}$ &
   $0.124\ ^{+\ 0.009}_{-\ 0.006}$ &  0.005 & $^{+0.008}_{-0.003}$
   & resum \\
$\epem$ [$\sigma_{\rm had}$]  & 34.0 &
 $0.146\ ^{+\ 0.031}_{-\ 0.026}$ &
   $0.123\ ^{+\ 0.021}_{-\ 0.019}$ & $^{+\ 0.021}_{-\ 0.019}
   $ & -- & NLO \\
\ul{$\epem$ [ev. shapes]} & 35.0 & $ 0.145\ ^{+\ 0.012}_{-\ 0.007}$ &
   $0.123\ ^{+\ 0.008}_{-\ 0.006}$ &  0.002 & $^{+0.008}_{-0.005}$
   & resum \\
\ul{$\epem$ [ev. shapes]} & 44.0 & $ 0.139\ ^{+\ 0.010}_{-\ 0.007}$ &
   $0.123\ ^{+\ 0.008}_{-\ 0.006}$ & 0.003 & $^{+0.007}_{-0.005}$
   & resum \\
$\epem$ [ev. shapes]  & 58.0 & $0.132\pm 0.008$ &
   $0.123 \pm 0.007$ & 0.003 & 0.007 & resum \\
%& & & & & & & \\
$\p\bar{\p} \rightarrow {\rm b\bar{b}X}$
    & 20.0 & $0.145\ ^{+\ 0.018\ }_{-\ 0.019\ }$ & $0.113 \pm 0.011$ 
    & $^{+\ 0.007}_{-\ 0.006}$ & $^{+\ 0.008}_{-\ 0.009}$ & NLO \\
${\rm p\bar{p},\ pp \rightarrow \gamma X}$  & 24.2 & $0.137
 \ ^{+\ 0.017}_{-\ 0.014}$ &
  $0.111\ ^{+\ 0.012\ }_{-\ 0.008\ }$ & 0.006 &
  $^{+\ 0.010}_{-\ 0.005}$ & NLO \\
%${\rm p\bar{p} \rightarrow W\ jets}$  & 80.6 & $0.123 \pm 0.025$ &
%  $0.121\pm 0.024$ & 0.017 & 0.016 & NLO \\
${\sigma (\rm p\bar{p} \rightarrow\  jets)}$  & 30 - 500 &  &
  $0.121\pm 0.009$ & 0.001 & 0.009 & NLO \\
\ul{$\epem$ [$\Gamma (\z0 \rightarrow {\rm had.})$]}
    & 91.2 & $0.122\pm 0.005$ & 
$0.122\pm 0.005$ &
   $ 0.004$ & $0.003$ & NNLO \\
$\epem$ [ev. shapes] &
    91.2 & $0.122 \pm 0.006$ & $0.122 \pm 0.006$ & $ 0.001$ & $
0.006$ & resum \\
%& & & & & & & \\
$\epem$ [ev. shapes]  & 133.0 & $0.111\pm 0.008$ &
   $0.117 \pm 0.008$ & 0.004 & 0.007 & resum \\
$\epem$ [ev. shapes]  & 161.0 & $0.105\pm 0.007$ &
   $0.114 \pm 0.008$ & 0.004 & 0.007 & resum \\
$\epem$ [ev. shapes]  & 172.0 & $0.102\pm 0.007$ &
   $0.111 \pm 0.008$ & 0.004 & 0.007 & resum \\
\ul{$\epem$ [ev. shapes]}  & 183.0 & $0.109\pm 0.005$ &
   $0.121 \pm 0.006$ & 0.002 & 0.006 & resum \\
\ul{$\epem$ [ev. shapes]} & 189.0 & $0.109\pm 0.006$ &
   $0.122 \pm 0.007$ & 0.003 & 0.006 & resum \\
%& & & & & & & \\
\hline
\end{tabular}
\end{center}
}
\end{table}
%%%%%%%%%%%%%%%%%%%%%%%%%%%%%%%%%%%%%% end table astab

\end{document}